\documentstyle[11pt,aaspp4]{article}
\input psfig.sty

\def\simgt{\lower.5ex\hbox{$\; \buildrel > \over \sim \;$}}
\def\simlt{\lower.5ex\hbox{$\; \buildrel < \over \sim \;$}}

\slugcomment{Submitted to the Astrophysical Journal Letters}
\lefthead{Gotthelf \& Kulkarni}
\righthead{Unusual X-ray Burst from M28}
\def\deg{\ifmmode^\circ\else$^\circ$\fi}

\def\simlt{\hbox{ \rlap{\raise 0.425ex\hbox{$<$}}\lower
0.65ex\hbox{$\sim$} }}
\def\ltorder{\hbox{ \rlap{\raise 0.425ex\hbox{$<$}}\lower
0.65ex\hbox{$\sim$}}}
\def\simgt{\hbox{ \rlap{\raise 0.425ex\hbox{$>$}}\lower
0.65ex\hbox{$\sim$} }}
\def\gtorder{\hbox{ \rlap{\raise 0.425ex\hbox{$>$}}\lower
0.65ex\hbox{$\sim$}}}
\def\sles{\lower2pt\hbox{$\buildrel {\scriptstyle <}
   \over {\scriptstyle\sim}$}}
\def\sgreat{\lower2pt\hbox{$\buildrel {\scriptstyle >}
   \over {\scriptstyle\sim}$}}

\def\qquad{\quad\quad}

\def\psr{PSR B1821$-$24}

\def\rxp{RX~J18245$-$2452P}
\def\rxe{RX~J18245$-$2452E}

\begin{document}

\title{An Unusual X-ray Burst from the Globular Cluster M28}

\author{E. V. Gotthelf\altaffilmark{1}}
\affil{Laboratory for High Energy Astrophysics \\ NASA/GSFC, Greenbelt, MD 20771}

\author{S. R. Kulkarni}

\affil{Division of Physics, Mathematics and Astronomy, 105-24 \\ California 
Institute of Technology, Pasadena, CA 91125, USA}

\altaffiltext{1}{Universities Space Research Association}

\begin{abstract}

We report the discovery of an unusual X-ray burst from the direction
of the Globular Cluster M28 using data acquired with the ASCA
Observatory.  The burst was recorded by all four ASCA telescopes and
displays a fast ($\simlt 70$ ms) rise followed by an exponential decay
($\tau = 7.5$ s) and a steady afterglow which lasts between 800 --
3250 s. The image of the burst is consistent with an ASCA point source
and is centered on quiescent X-ray emission from the core of M28. The
burst temporal profile is similar to Type I bursts emitted by
accreting neutron stars of low mass X-ray binaries (LMXB). We argue
that the burst arises from an LMXB that is located in the core of M28.
The burst is unique in two ways: it is intrinsically sub-luminous,
$\sim 0.02L_E$ and more importantly, originates from a source whose
quiescent luminosity is fainter than that of the known cluster
bursters by three orders of magnitude.  We suggest that this burst is
from a highly magnetized neutron star accreting at a low rate. These
accreting systems may account for the mysterious low luminosity X-ray
sources in globular clusters.
\end{abstract}



\section{INTRODUCTION}

The low mass X-ray binaries (LMXB) are the brightest X-ray sources in
the sky. About 124 are known of which $\sim$ 40 show Type I bursts
(Lewin, van Paradijs, \&\ Taam 1993). These bursts are characterized
by a sharp rise time and a spectrum which becomes softer as the burst
evolves. The bursts have been interpreted as cooling of a thin layer
of material on the surface of neutron stars that is heated by unstable
nuclear fusion within this layer (Woosley \&\ Taam 1976; Maraschi \&\
Cavaliere 1977).  The peak luminosity appears to have a small
dispersion, less than a factor of 10 with a maximum value equal to the
Eddington luminosity, $L_E\sim 10^{38}$ erg s$^{-1}$.  A
disproportionate number of LMXBs are found in the Galactic globular
clusters and all these have exhibited a Type I burst (Dotani et al.
1990).

A day long observation of the globular cluster M28 was conducted with
the ASCA X-ray Observatory (Tanaka et al. 1994) on March 29, 1995 UT.
The original motivation was to search for X-ray pulsations from the
known 3-ms radio pulsar \psr\ (Lyne et al. 1987). These pulsations were
subsequently detected (Saito et al. 1997).  Recently, during the
course of verification of a timing package, we analyzed this data. In
the process we discovered an unusual X-ray flash, the analysis of
which is reported in this Letter.

\section{OBSERVATIONS}

The ASCA satellite consists of four co-aligned telescopes, each of
which is a conical foil mirror coupled to one of two types of focal
plane sensors, the two Solid State Imaging Spectrometers (SIS--0 \&
SIS--1) or the two Gas Imaging Spectrometers (GIS--2 \& GIS--3). The
SIS detectors are sensitive to photons in the $0.4-10.0$ keV energy
band with a nominal spectral resolution of 2\% at 6 keV, scaling as
$\sim E^{-1/2}$.  The angular resolution of the SIS is $\sim
1^{\prime}$, limited by the mirror optics.  The SIS data were acquired
in the ``4-CCD'' mode, using 16 s integrations, with the $22'\times 22'$ 
SIS field-of-view centered on the core of the M28 cluster.

The GIS detectors have a relatively modest spectral resolution of 8\%
at 6 keV, in comparison to the SIS, with a larger effective area at
higher energies. The temporal resolution is also higher, 61 $\mu$s or
0.5 ms, depending on the mode that was selected. The images formed by
the GIS sufferes from an energy dependent blur which results in an
effective angular resolution of 3$^{\prime}$. The GIS data were edited
using the standard ``REV1'' processing criteria to exclude periods of
high background contamination. The SIS data were time filtered using
the GIS good time intervals to directly compare the light curves from
both instruments. The final effective exposure time is $\sim 37$ ks
per detector.

\section{RESULTS}

The ASCA image of the M28 region obtained using the entire data set
contains a single unresolved source, AX J18245$-$2451. This source is
located at the light centroid of the globular cluster M28 and has been
studied in detail by Saito et al. (1997). We constructed light curves
using photons from a $12^{\prime}$ diameter aperture centered on AX
J18245$-$2451 and found a flash, lasting tens of seconds, followed by
an afterglow. The profile of the burst is essentially identical in all
instruments (Figure 1). The background light curves from each detector
show no appreciable rise in the overall count-rate during the flash or
the afterglow. The summed background light-curve is shown in Fig. 1 and 
contains photons from a concentric annulus of radius
$12^{\prime} - 20^{\prime}$ (GIS) and $> 12^{\prime}$ (SIS).

The background particle flux rate during the burst interval, as
measured by the GIS radiation belt monitor, shows no unusual
activity during this period. A careful study of the GIS scaler
(discriminator) rates, which record the number of detections in each
GIS before any detailed processing is done, also show no abnormal
behavior. The scaler rates during the burst are at no time high enough
to warrant dead-time correction. As expected, the burst count-rates
measured by each detector are not the same. This is because the
four mirrors are not quite co-aligned, and a source will fall at
different off-axis angles for each mirror and thus suffer a varying
amount of mirror vignetting (geometric shadowing). A second
reason is that the relative detector efficiency is a strong function
of position.  The combined effect of the two is typically $\sim
30\%$. Indeed, we find that the relative burst fluence measured from
each detector is consistent with that expected for a cosmic (i.e. a
distant) source imaged through the telescopes; the relative
count-rates for the burst are identical to those measured for the quiecent
emission, within counting statistics and spectral differences.

Figure 2 shows the combined SIS (SIS--0 + SIS--1) images and
spectra of AX J18245$-$2451 obtained during the time intervals defined
in Table I: pre-burst, flash, afterglow, and post-burst.  The images
are centered on AX J18245$-$2451 and display the $22^{\prime}\times 
22^{\prime}$ field-of-view of the SIS. The burst and afterglow photons 
form an image which, within the photon counting errors, is a point 
source with the same pixel centroid as AX J18245$-$2451; the 95\% confidence 
radius of the burst localization is $40^{\prime\prime}$ (see Gotthelf 1996 
for details on measurement accuracy with the SIS).

The facts presented so far -- (1) the detection of the burst in all
four detectors, (2) the absence of any abnormal behavior in the
background rate, (3) the consistency of the image of the burst in each
detector, including the variation in fluence expected from telescope
vignetting -- lead us to conclude that the burst is cosmic in origin.
This conclusion motivated us to carry out a detailed analysis of the
burst.

For each of the four sub-time intervals specified in Table 1 we created a
spectrum from photons extracted from a $12^\prime$ diameter circular
region centered on the image peak. These spectra are displayed in
Fig. 2. To compare spectra from different time intervals we have
normalized them by duration and plotted them auto-scaled to the peak
value. The burst rate is almost 2 orders of magnitude higher than the
quiescent rate.  Even without any sophisticated analysis we note that
the spectrum of the burst is distinctly different from that of the
quiescent emission. The results discussed here are identically
reproduced by the GIS data.

As with the SIS data, GIS photons within a 12$^{\prime}$ diameter
circular region centered on AX J18245$-$2451 were extracted. The burst
profile derived from the combined GIS data is shown in Fig. 3.  The
rise of the burst is unresolved. A photon-by-photon arrival time
analysis of the high time resolution GIS data yields a turn-on time of
$<$ 70 ms. The burst decay signal is well represented by an
exponential model, $I(t) = I_p e^{-t/\tau} + I_aS(t) + I_q.$, where
$S(t)$ is the step function, $S(t)=0, t<0$ and $S(t)=1, t>0$ and the
subscripts $p, a, \ \&\ q$ refer to the peak, afterglow and quiescent
(pre-burst) counting rates. Fitting the data to this functional form
we obtain $I_{p} = 28.0$ cps, $I_{a} = 0.20$ cps, $I_{q}=0.08$ cps and
$\tau = 7.5$ s; this model is displayed in Fig. 3 (solid line). The
start time of the first significant burst emission measured by the two
instruments is consistent with counting statistics ($\sim$ inverse of
the count rate). The estimated uncertainty in the onset of the burst
is about 40 ms. The afterglow increases the quiescent (pre-burst)
count rate $\sim 2.5$ times and lasts between 800 s to 3250 s; the gap
in the data is due to Earth block.

Comparison of the burst profile in the hard ($>2.0$ keV) and soft-band
($<2.0$ keV; see Figure 3) shows some indication of spectral evolution
from the initial soft impulse. Specifically, the spectrum is seen to
soften (i.e. more photons at lower energies relative to higher
energies) as the burst evolves.

Spectral fits to the mean burst spectrum obtained from the two GISs
are presented in Table 2.  The background is a negligible contribution
during this interval. We fit several standard models to the data, all
of which produced reasonable fits. This is not surprising
given the small number of counts. It is noteworthy that all three
models require a best-fit $N_H$ value larger than that
obtained by Saito et al. (1997) from analysis of the 3-ms pulsar data.
Saito et al.'s $N_H$ value of $2.8\times 10^{21}$ cm$^{-2}$ is
consistent with other measures of the interstellar absorption expected
to the cluster M28. Thus we conclude that there is an additional
absorption internal to the bursting source.

\section{DISCUSSION}

The duration of the flash is so short that we can safely exclude
stellar flares (Greiner, Duerbeck \&\ Gershberg 1994) from an active
star such as UV Ceti or a T Tauri (young) stars. Our flash was not
coincident in time with any known gamma ray burst, making it unlikely
to be an X-ray precursor.  This flash bears some similarity to the
flashes reported from analysis of the database of the X-ray satellite
EINSTEIN (Gotthelf, Hamilton \&\ Helfand 1996). But the sky
distribution of the EINSTEIN flashes does not suggest a globular
cluster origin.  We note that both the temporal and spectral
properties of our flash are similar to those of Type I bursts.
Furthermore, the flash is coincident with the core of a globular
cluster and clusters are known to be relatively abundant in LMXBs. To
our knowledge all LMXB's in clusters have exhibited a Type I burst
(Dotani et al. 1990). Thus we adopt the simplest hypothesis, viz.,
that this is a Type I burst originating in the globular cluster M28.

Using the mean spectrum for the flash (Table 2) we obtain a peak flux
of $1.4\times 10^{-9}$ erg cm$^{-2}$ s$^{-1}$ which translates to an
(assumed) isotropic luminosity of $L_B\sim 4.3\times 10^{36}$ erg
s$^{-1}$; the distance to M28 is assumed to be 5.1 kpc (Rees \&\
Cudworth 1991).  The total energy in the burst is $E_B\sim 3.1\times
10^{37}$ ergs.  Almost all Type I bursts reported in the literature
have peak luminosities $fL_E$ with $f$ ranging from more than 0.1 to 1
where $L_E \sim 2\times 10^{38}$ erg s$^{-1}$ is the Eddington
luminosity for a neutron star.  Our burst appears to have $f\sim
0.02$. This may appear to be in conflict with our conclusion that the
burst under discussion is a Type I burst.  However, below, we argue
that this is not a fatal objection.

Type I bursts are understood to be thermo-nuclear instabilities of the
accreted matter on neutron stars.  Once a critical column density,
$C_{\rm cr}$, is reached nuclear fusion produces far more heat
than can be radiated away.  This results in the layer getting heated till
expansion sets in. The resulting cooling accounts for the
softening of the spectrum as the burst progresses.  At any given point
on the surface of the neutron star, the critical column density is a
function of the composition (including metallicity) of the accreted
gas, the core temperature of the neutron star and the {\it local}
accretion rate (Taam \&\ Picklum 1978; Joss \&\ Li 1980; Fujimoto,
Hanawa \&\ Miyaji 1981).  $C_{\rm cr}$ (Fujimoto, Hanawa \&\ Miyaji
1981; Bildsten \&\ Brown 1997) varies from $2\times 10^8$ gm cm$^{-2}$
to $2\times 10^7$ gm cm$^{-2}$ as the core temperature varies from
$10^{7.5}$ K to $10^9$K. 

Strong magnetic fields force the accretion to polar cap regions,
thereby changing the local accretion rate. This fact explains why
accreting X-ray pulsars (which are highly magnetized neutron stars) are
not seen to burst and likewise why bursters do not pulse.  For a given
source, $C_{\rm cr}$ is not expected to be highly time dependent.  For
values of $C_{\rm cr}$ applicable to most bursters, the observed range
in $f$ translates to an inference that the bursts are from a
substantial portion, 10\% to 100\%, of the surface of the neutron star.
This inference fits in well with our ideas that LMXBs are magnetically
weak neutron stars and thus there is little impediment to prevent the
spread of accreted material.

There is an alternative explanation for sub-luminous Type I bursts.
Garcia \&\ Grindlay (1987) noted that a Type I burst from
4U2129+47 had an even lower $f$ value, $\sim 0.002$.  They
persuasively argue that 4U2129+47 is a LMXB whose thick accretion
disk obscures the neutron star from our sight.  From our vantage point
we only see a fraction of the quiescent and the burst emission that is
scattered by the electrons in the corona.  This accounts for the small
value of $f$.  This model also explains two other peculiarities of the
4U2129+47 burst, a longer rise time (6 s instead of the usual $<$1 s)
and the higher effective temperature, $\sim 3$ keV instead of $\simlt
2$ keV. Scattering allows multi-path propagation which then accounts
for the longer rise time.  Inverse Compton scattering of photons by
the hot electrons in the corona also boosts the energy of the incident
photon.

However, the flash reported here has a soft spectrum and a short rise
time and appears to be no different from a typical Type I burst.  Thus
we conclude that our flash is a Type I burst but intrinsically
sub-luminous.  If so, we are forced to the conclusion that material
from only a small area of the neutron star, between 0.1\% to 1\%, was
involved in this burst, depending on the value of $C_{\rm cr}$ we
adopt.  For a low luminosity source, $C_{\rm cr}$ is higher and thus
we prefer the smaller value for the area.  The accreted material is
probably confined by a strong magnetic field. With such a small area
for the burst we should expect to see very deep modulation due to
rotation.  The lack of a significant modulation in the burst profile
indicates that we are either looking down at the magnetic pole or that
the neutron star's rotation period is much greater than 10 s.

In the Type I burster model, the burster must have a certain quiescent
X-ray luminosity. The quiescent emission is generated from accretion
of matter into the deep potential well of the neutron star.  This gas
subsequently undergoes nuclear fusion. Therefore, one expects the
ratio of the quiescent luminosity to the mean burst luminosity to be
the ratio of the potential energy of one gram of matter to the nuclear
energy liberated from the same gram or about 25 (H-burning) or 100
(He-burning).  Thus the mean time between bursts, $T$,
is expected to satisfy the following relation:  $T \simgt
\alpha E_B/L_Q$.  Here $E_B$ is the integrated energy of the burst and
$L_Q$ is the luminosity due to accretion i.e. the luminosity of the
source in quiescence.

From observation of the M28 region with the ROSAT High Resolution
Imager (HRI) only two X-ray sources are found in the entire HRI
field-of-view, the 3-ms radio pulsar \rxp\ and \rxe\ (Danner et al.
1997). These are separated by $10^{\prime\prime}$ in the HRI images.
However, at the resolution of the ASCA mirrors these two sources blend
together and thus our ASCA source AX J18245$-$2451 is the sum of the
two ROSAT sources.  Danner et al. (1997) speculate that \rxe\ is
either a collection of low luminosity point sources or a synchrotron
nebula. An upper limit to the quiescent flux of the burster is thus a
fraction of the luminosity of \rxe, say $L_Q\simlt 10^{33}$ erg
s$^{-1}$.  Given that the energy of our burst is $E_B\sim 3.1\times
10^{37}$ erg we find $T\sim 0.1 (\alpha/100)$ yr.  This estimate of
$T$ is consistent with our lack of detections of any other burst from
M28 in archival X-ray data from the ROSAT or EINSTEIN missions.

\section{SUMMARY}

We have discovered a Type I burst which is unique in two ways: (1) it
is sub-Eddington by a factor of 100 and more importantly (2) the
quiescent luminosity of the source is three orders of magnitude
smaller than that of all previously known bursters. We propose that
these two observations can be explained by a model of a highly
magnetized neutron star accreting at a low rate. The low accretion
rate results in an Alv\`en radius that is large and thus an
equilibrium spin period that is concomitantly long (Bidlsten \&\ Brown
1997).  This provides a natural explanation for the observed lack of
any modulation in the burst profile despite the fact that the burst
supposedly arises from a very small region.

Radio pulsars in globular clusters have a range of magnetic field
strengths; see review by Kulkarni \&\ Anderson (1996). These authors
have argued that this observation can be explained by assuming that
all cluster neutron stars are born with a distribution of magnetic
field strengths similar to those in the disk (i.e. a typical dipole
field strength of $10^{12}$ G). Depending on the accretion history a
range of field strengths are then obtained. However, do note that
Lyne, Manchester and D'Amico (1996) disagree with this conclusion.
Leaving aside this dissenting view we note that in the picture of
Kulkarni \&\ Anderson highly magnetized neutron star accreting matter
at a low rate are expected to be quite common. Thus a natural
explanation can be found in this framework for the type of LMXB we
have hypothesized as the the source of the burst discussed in this
paper.

A natural consequence of this suggestion is that similar flashes
should also be seen towards other clusters. Why have not such
anomalous bursts been noticed before?  We argue that previous
non-imaging missions missed these bursts because of their limited
dynamic range in sensitivity.  Imaging telescopes have the sensitivity
to detect such bursts.  Unfortunately, the frequency of such flashes
is low, a dozen per year per rich cluster. Discovery of another such
burst from M28 would require two months of dedicated observations by
the ASCA satellite! Thus it will not be easy to verify this
prediction. We eagerly look forward to the deployment of large
field-of-view imaging monitors such as that proposed by Priedhorsky et
al. (1996).

Finally, we note that this class of bursts offers new insights into
the physics of thermonuclear instabilities.  For example, the fact
that the quiescent luminosity is low enabled us to clearly recognize
the afterglow. This phenomenon in itself is of interest to models of
Type I bursts. The afterglow could arise from extended burning
(Fushiki et al. 1992) or slow propagation of the burning front
(Bildsten 1995) or simply the emission from a transiently heated
surface. The first two reasons have been advocated for conventional
Type I bursts but the latter is a new twist. It is important to note
that the surface of the neutron stars in these systems is expected to
be cooler than those of the brighter bursters, both because of a very
small burst rate as well as a small quiescent luminosity.  As remarked
earlier, the value of the surface temperature (core temperature) is an
important parameter in the theory of Type I bursts.  This opens the
possibility that such bursts offer tests of the model over a range of
previously unexplored parameter space.

\bigskip

\noindent
{ Acknowledgments: EVG's research is supported by NASA. SRK's research
is supported by NASA and NSF. SRK thanks L. Bildsten and W. Lewin for
education on the topic of bursters.  We thank G. Hasinger, D. Helfand,
E. S. Phinney, J. Swank, and F. Verbunt for helpful discussions and
K. Ebisawa for useful discussion concerning the GIS scaler data. This
research has made use of data obtained through the HEASARC Online
Service, provided by the NASA/GSFC.}

\vfill\eject


\hfil\vbox{\halign{#\hfil & #\hfil & \hfil #\hfil  & \hfil #\hfil & \hfil #\hfil \hfil \hfil\tabskip=0pt\cr\multispan5\hfil{Table 1}\hfil\cr
\multispan5\hfil{M28 Observation Log and Burst Time Intervals}\hfil\cr
\noalign{\vskip 1em\hrule\vskip 2pt\hrule\vskip 1em}
Interval & Start Time & Exposure$^a$     & Counts$^b$ &Rates\cr
         &    (UT)    & net(total) (sec) &            &(cps)\cr
\noalign{\vskip 1em\hrule\vskip 2em}
Observation  & 28/03/95 17:23:36  &  74,220(207,328) & 6,654 &0.09\cr
Pre-burst    & 28/03/95 17:23:36  &  62,581(181,664) &  5137 &0.08\cr
Burst        & 29/03/95 18:37:33  &       90(90)     &   212 &2.21$^c$\cr
Afterglow    & 29/03/95 18:38:18  &     1,440(1,528) &   400 &0.28\cr
Post-burst   & 29/03/95 18:51:02  &   10,109(24,036) &   905 &0.09\cr
\noalign{\vskip 1em\hrule\vskip 2em}
\multispan5{ $^a$ Screened (i.e. Good Time Interval, GTI) and total exposure time\hfil}\cr
\multispan5{   from both GIS sensors.}\hfil\cr
\multispan5{ $^b$ Total photons falling within a 6 arcmin radius aperture centered}\hfil\cr
\multispan5{   on the X-ray source (both GIS sensors).}\hfil\cr
\multispan5{ $^c$ Peak rate is 28 counts per sec.}\hfil\cr
}\hfil}

\hfil\vbox{\halign{#\hfil & \hfil#\hfil & \hfil#\hfil & \hfil#\hfil & \hfil#\hfil\tabskip=0pt\cr\multispan5\hfil{Table 2}\hfil\cr
\multispan5\hfil{M28 Burst GIS Spectral Fits$^a$}\hfil\cr
\noalign{\vskip 1em\hrule\vskip 2pt\hrule\vskip 1em}
    Model & Continuum        & N$_H^I$ $^b$              &    Fluence$^c$  & $\chi_{\nu}^2$ ($\nu$)\cr
          & $\Gamma$ or  kT  & ($\times 10^{22}$ cm$^2$) & (erg cm$^{-2}$) &                       \cr
\noalign{\vskip 1em\hrule\vskip 2em}
Power Law & $3.7_{-0.9}^{+1.2}$ & $2.2_{-1.0}^{+1.2}$  & $1.1  \times 10^{-8}$ & 0.9 (12)   \cr
Blackbody & $0.6_{-0.1}^{+0.1}$ & $0.35_{-0.3}^{+0.9}$ & $7.7 \times 10^{-9}$ & 1.3 (12)  \cr
R--S      & $1.6_{-0.5}^{+1.0}$ & $1.3_{-0.9}^{+2.6}$  & $1.1  \times 10^{-8}$ & 1.1 (12)  \cr
\noalign{\vskip 1em\hrule\vskip 2em}
\multispan5{$^a$ GIS-2 \& GIS-3 photons collected using a 6 arcmin aperture \hfil}\cr 
\multispan5{  centered on the source with no background\hfil}\cr
\multispan5{  subtraction (assumed negligible). Spectral fits are in the \hfil}\cr
\multispan5{  $1.0-10.0$ keV range. Errors are 90\% confidence region for \hfil}\cr
\multispan5{  a single interesting parameter.\hfil}\cr
\multispan5{$^b$ Intrinsic absorption, after allowing for an interstellar\hfil}\cr
\multispan5{  absorption of $2.9 \times 10^{21}$ cm$^2$ (from Saito et al. 1997) \hfil}\cr
\multispan5{$^c$ The unabsorbed fluence at the front of the telescope, integrated\hfil}\cr
\multispan5{   during the 45 sec burst interval.\hfil}\cr
}\hfil}

\vfill\eject

\begin{figure}

\centerline{\psfig{figure=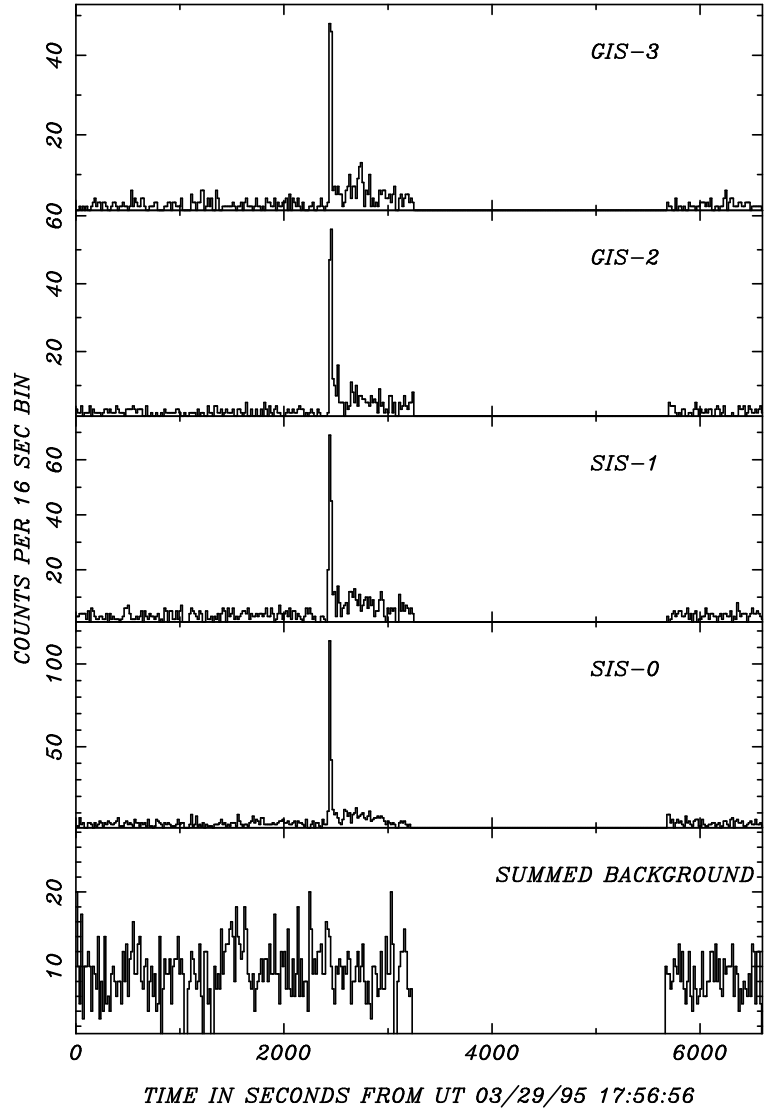,height=6.0in,angle=00.0,bbllx=25bp,bblly=25bp,bburx=400bp,bbury=525bp,clip=}}

\caption{ The burst from M28 was observed by all four telescopes
aboard the ASCA Observatory. The burst profile is essentially identical 
in all instruments with a bright peak and a distinct afterglow signature. 
These light curves are binned in 16 s steps. The amplitude of the burst 
(as defined by the height of the peak bin) is determined by the large, 
undersampled, binning and is somewhat arbitrary. No appreciable rise in the 
background rate accompanied the burst; the burst is coincident with the 
quiescent flux from the core of M28. The data gap between the interval 
3.2 and 5.6 ksec is due to SAA passage.}

\end{figure}

\begin{figure}

\centerline{\psfig{figure=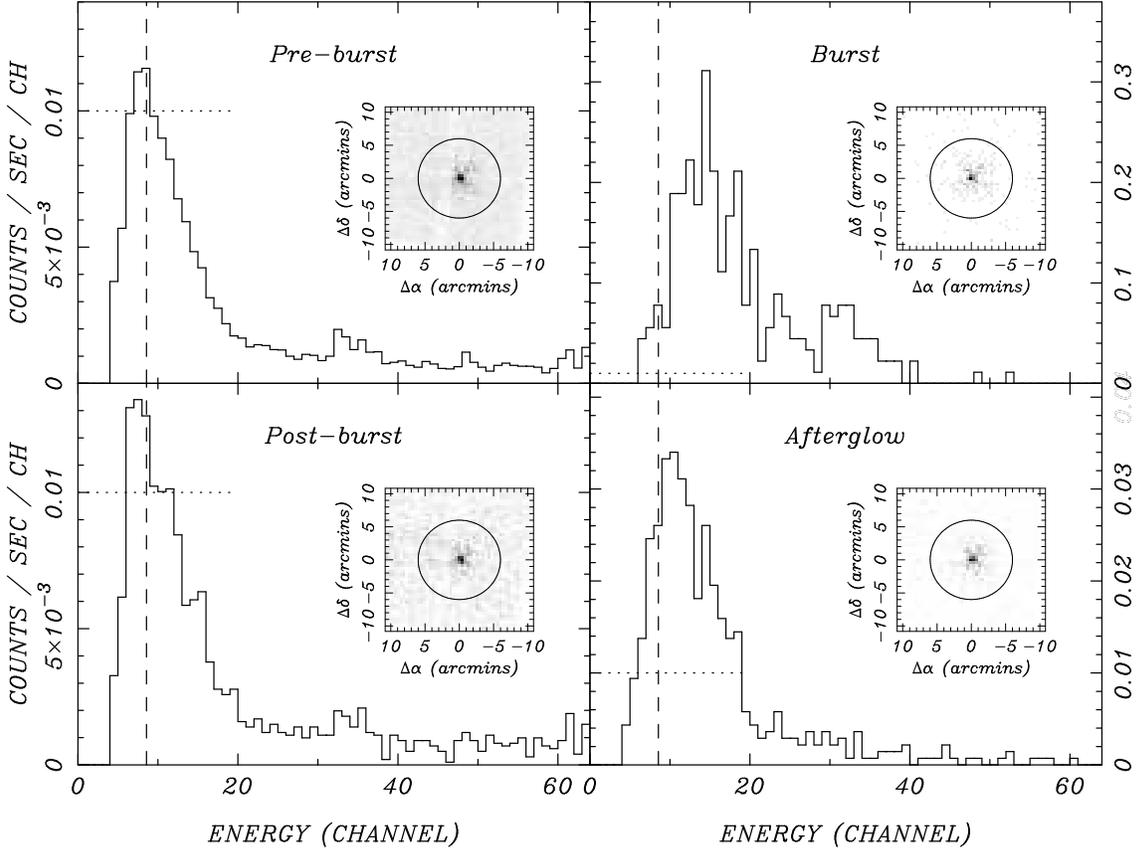,height=5.0in,angle=270.0}}

\caption{ The combined SIS (SIS--0 + SIS--1) images and
spectra of AX J18245$-$2451 obtain during the time intervals defined in Table I:
pre-burst (top left), flash (top right), afterglow (bottom right), and
post-burst (bottom left).  The images are centered on the (same)
measured centroid of AX J18245$-$2451 and display the entire 22' X 22' field-of-view of the
SIS. The burst and afterglow images are consistent with a celestial point
source (of equivalent counts) focused on the detectors. The $12^\prime$
diameter circle denotes the extraction region for the spectra plotted
here, which are gain corrected and shown rebined in units of $\sim$ 125 eV per bin.  
The dashed vertical line indicates the energy channel equivalent of 1 keV.  
The dashed horizontal line denotes a fiducial count rate of 0.01 cps per bin; the burst rate is
almost 2 orders of magnitude higher then the quiescent rate.  These results 
are perfectly reproduced by the GIS data (not shown).
}

\end{figure}

\begin{figure}

\centerline{\psfig{figure=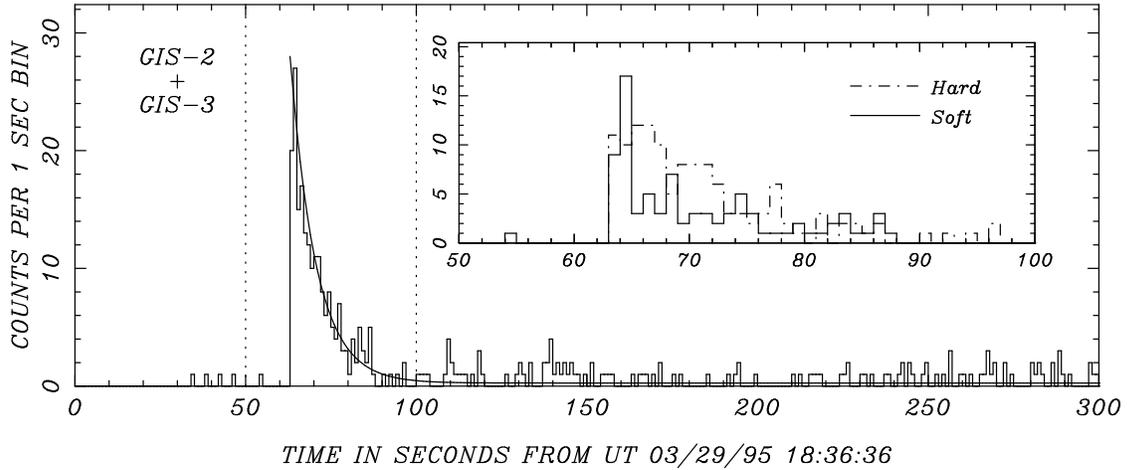,height=5.0in,angle=270.0,bbllx=250bp,bburx=580bp,clip=}}

\caption{The combined GIS--2 and GIS--3 light curve binned
in 1 sec steps and centered on the prominent burst interval (histogram). The
burst risetime is unresolved and is limited by counting statistics to $<
70 $ msec.  The decay signal is well represented by an exponential model 
(solid line) with a e-fold time of $\tau = 7.5$ s. (Insert) The
spectrally resolved light curves.  The dash curves includes only
photons with channel equivalent energy above 2 keV, whereas the solid
histogram is for the remaining, soft photons. There is some indication
of spectral evolution (see text).
}

\end{figure}

\end{document}